\documentclass{article}
\usepackage[T1]{fontenc} 
\usepackage[utf8]{inputenc} 
\usepackage{ismir,amsmath,cite,url}
\usepackage{graphicx}
\usepackage{color}
\usepackage{booktabs}
\usepackage{tabularray}
\usepackage{multirow}
\usepackage{amsmath}
\usepackage{amssymb}
\usepackage{array}
\usepackage{makecell}
\usepackage{lineno}
\usepackage{xcolor}
\usepackage{microtype}

\title{Cluster and Separate: a GNN Approach to \\ Voice and Staff Prediction for Score Engraving}

\multauthor{Francesco Foscarin$^{*1,2}\thanks{* Equal contribution.}$ \hspace{0.3cm} Emmanouil Karystinaios$^{*1}$ \hspace{0.3cm} Eita Nakamura$^{3}$ \hspace{0.3cm} Gerhard Widmer$^{1,2}$}
{$^1$ Johannes Kepler University, Linz, Austria\\
$^2$ LIT AI Lab, Linz Institute of Technology, Austria\\
$^3$ Kyushu University, Japan\\ {\tt\small firstname.lastname@jku.at}}

\def\authorname{F. Foscarin, E. Karystinaios, E. Nakamura and G. Widmer}

\newcommand{\V}{V}

\newcommand{\Ein}{\ensuremath{E_{in}}}

\begin{document}

\maketitle

\begin{abstract}
This paper approaches the problem of separating the notes from a quantised symbolic music piece (e.g., a MIDI file) into multiple voices and staves. This is a fundamental part of the larger task of music score engraving (or score typesetting), which aims to produce readable musical scores for human performers. We focus on piano music and support homophonic voices, i.e., voices that can contain chords, and cross-staff voices, which are notably difficult tasks that have often been overlooked in previous research. We propose an end-to-end system based on graph neural networks that clusters notes that belong to the same chord and connects them with edges if they are part of a voice. Our results show clear and consistent improvements over a previous approach on two datasets of different styles. To aid the qualitative analysis of our results, we support the export in symbolic music formats and provide a direct visualisation of our outputs graph over the musical score. All code and pre-trained models are available at \url{https://github.com/CPJKU/piano_svsep}.

\end{abstract}

\section{Introduction}\label{sec:introduction}



The musical score is an important tool for musicians due to its ability to convey musical information in a compact graphical form. Compared to other music representations that may be easier to define and process for machines, for example, MIDI files, the musical score is characterized by how efficiently trained musicians can read it. 

\begin{figure}
    \centering
    \includegraphics[width=0.8\columnwidth]{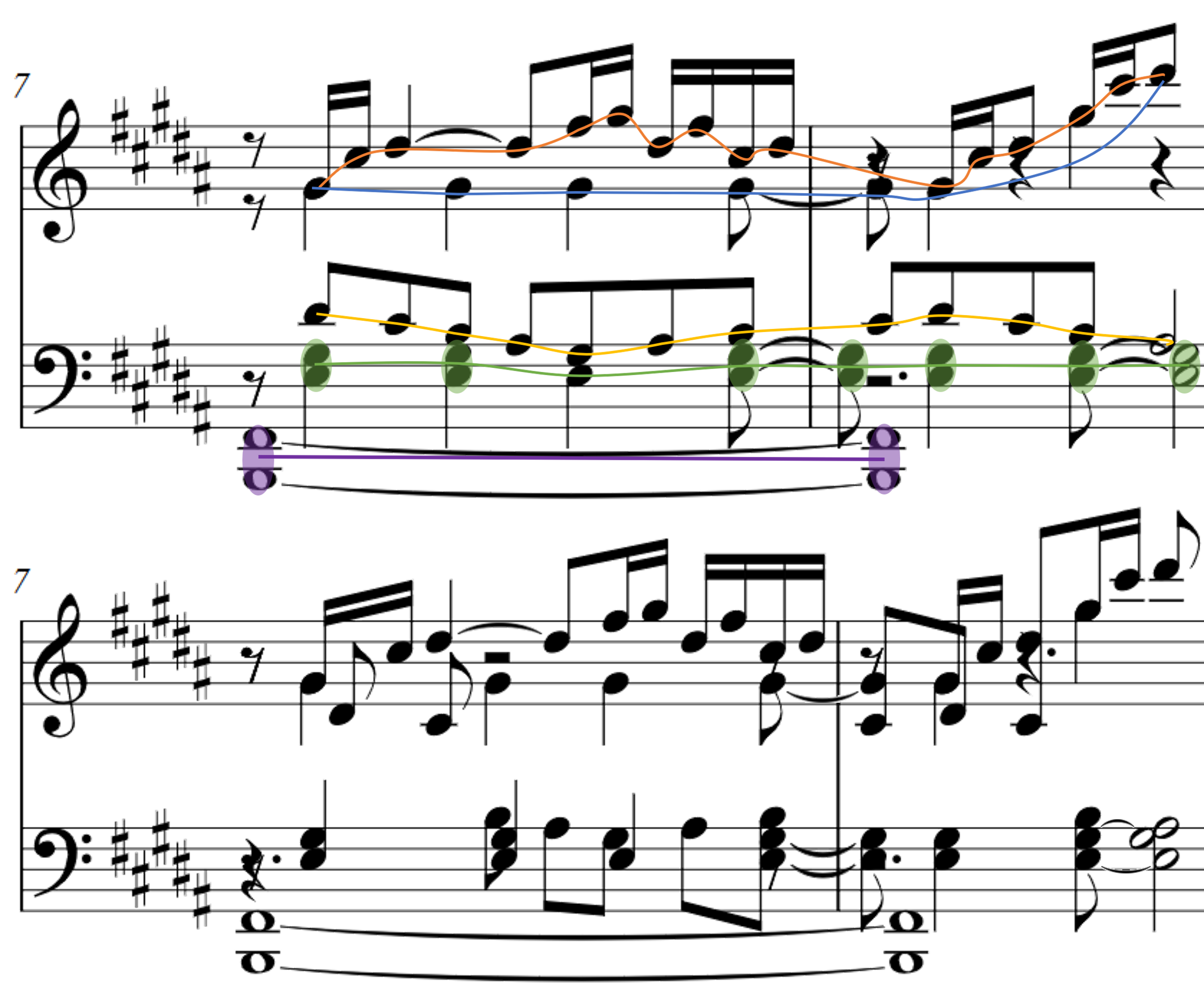}
    \caption{Comparing different voice/staff assignments for two bars from C. Debussy's Estampes - Pagodes. (top) original; voices can be inferred from the beam grouping and (horizontal lines connecting notes), rests, and stem sharing, and are colored for clarity. (bottom) hard-to-read rendition with voice and staff assigned according to heuristics we propose as a baseline.
    }
    \label{fig:random_voices}
\end{figure}

An important factor that affects the readability of a musical score for instruments that can produce more than one note simultaneously, is the separation of notes into different \textit{voices} (see Figure~\ref{fig:random_voices}). This division may follow what a listener perceives as independent auditory streams~\cite{cambouropoulos2008voice}, which can also be reflected in a clearer visual rendition
of a musical score~\cite{gould2016behind}. A similar point can be made for the division into multiple staves (generally 2) for instruments with a large pitch range, such as piano, organ, harp, or marimba. We will consider in this paper piano music.

The term voice is frequently used to describe a sequence of musical notes that do not overlap, which we call a \textit{monophonic voice}. However, this definition may be insufficient when considering polyphonic instruments.
Voices could contain \textit{ chords}, which are groups of \textit{synchronous notes} (i.e., notes with the same onset and offset) and are perceived as a single entity~\cite{makris2016visa3}. We name a voice that can contain chords a \textit{homophonic voice}. Note that partially overlapping notes cannot be part of a homophonic voice.

Music encoded in MIDI (or similar) formats, even when containing quantized notes, time signature, or bar information, often does not contain voice and staff information. The same can be said for the output of music generation~\cite{10.1145/3394171.3413671}, transcription~\cite{shibata2021non}, or arranging~\cite{terao2023neural} systems. 
Therefore, such music cannot be effectively converted into a musical score, to be efficiently read and played by human musicians.\footnote{Voice and staff separation are only two of the multiple elements, such as pitch spelling, rhythmic grouping, and tuplet creations, which need to be targeted by a score engraving system, but we will only focus on the former two in this paper.} The tasks of producing voice and staff information from unstructured symbolic music input are called \textit{voice separation} (or voice segregation in some papers~\cite{makris2016visa3}) and \textit{staff separation}, respectively.

Most of the existing approaches to voice separation have focused only on music with monophonic voices~\cite{chew2004separating,duane2009streaming,gray2016neural,mcleod2016hmm,hsiao2021learning,karystinaios2023voice}, which is not sufficient for our goal of engraving\footnote{``score engraving'' and ``score typesetting'' are used interchangingly.} piano music. 
The task of homophonic voice separation is much harder to solve: the presence of chords within voices makes the space of solutions grow much bigger; and the choice of the ``true voice separation'' can be ambiguous, with multiple valid alternatives among which experts may disagree.

The existing approaches to homophonic voice separation can be divided into two groups: the first~\cite{cambouropoulos2008voice,kilian2002voice,makris2016visa3,shibata2021non} use dynamic programming algorithms based on a set of heuristics, which makes for systems that are controllable and interpretable, but also hard to develop and tune. Such systems are often prone to fail on exceptions and corner cases that are present in musical pieces. 
The second group of approaches~\cite{suzuki2021score,hiramatsu2021joint,liu2022performance,ijcai2023p652} applies deep learning models to predict a voice label for each note. Such an approach creates two fundamental issues: i) the necessity of setting a maximum number of voice labels, and ii) a (highly) unbalanced ratio of occurrence of some voice labels. 
Moreover, all these approaches assume
that a voice cannot move between the two staves, which is not true for complex piano pieces.


In this work, we propose a novel system for homophonic voice separation that can efficiently and effectively assign notes to voices and staves for polyphonic music engraving.
Efficiency is ensured by a graph neural network (GNN) encoder, which can create input embeddings with a relatively small number of parameters. Effectiveness is targeted by approaching voice prediction not as a note labeling, but as an edge prediction problem~\cite{karystinaios2023voice}, which solves the maximum voice number and the label imbalance problems presented above. 
Our system predicts staff and voice separately and does not make any assumption on the number of voices; therefore it can deal with cross-staff voices and complex corner cases.
We avoid the problem of ground truth ambiguity since we focus specifically on voice separation for musical score engraving, therefore we can extract the (unique) ground truth directly from digitized musical scores.

We evaluate our system on two piano datasets of different difficulty levels, one containing popular, the other classical music. A comparison with a baseline and the approach of Shibata et al.~\cite{shibata2021non} shows a consistent improvement in performance on both datasets.
Finally, we develop a visualization tool to display the input and output of our system directly on the musical score, and discuss some predictions and comments on homophonic voice separation.

\section{Related Work}\label{sec:related}
The most influential work for this paper is the monophonic voice separation system by Karystinaios et al.~\cite{karystinaios2023voice}. Similarly, we consider voice separation a edge prediction task and use a similar score-to-graph routine and the same GNN encoder. Differently from that work, we consider homophonic voices and staves and, therefore, we extend the model formulation, the deep learning architecture, and the postprocessing routine to deal with this information. 

Shibata et al.~\cite{shibata2021non} developed a voice and staff separation technique applied after music transcription to quantized MIDI files. It works in two stages: first, an HMM separates the notes of the two hands (which will then be used as staff), and then a dynamic programming algorithm that maximizes the adherence to a set of heuristics is applied to separate voices in the two hands independently. We compare against this method since it is the most recent approach focusing specifically on homophonic voice separation.

There are some approaches based on neural networks~\cite{hiramatsu2021joint,liu2022performance, suzuki2021score,ijcai2023p652}, but they never perform this task in isolation, but rather in combination with other tasks such as symbolic music transcription, full scorification, and automatic arrangement. This means that they can only train on a much smaller dataset and a comparison would not be fair. 

All the approaches mentioned before, except~\cite{karystinaios2023voice}, perform voice separation as a label prediction task, which is problematic, as discussed in the introduction, due to the label imbalance and choice of the maximum number of voices. The former is particularly problematic for the neural network approaches.
\section{Methodology}\label{sec:Methodology}
Our system inputs data in the form of a set of quantized notes (e.g., coming from a quantized MIDI or a digitized musical score), each characterized by pitch, onset, and offset.
This information is modeled as a graph, which we call \textit{input graph}, and then passed through a GNN model to predict an \textit{output graph} containing information about voices, staves, and chord groupings. We remind the reader that in our `homophonic voice' scenario, chords are groups of synchronous notes that belong to the same voice.

\subsection{Input Graph}\label{subsec:input_graph}

From the set of quantized notes representing a musical piece we create a directed heterogeneous graph~\cite{hamilton2017representation} $G_{in} = (\V, \Ein, \mathcal{R}_{in})$ where each node $v \in V$ corresponds to one and only one note, and the edges $e \in \Ein$ of type $r \in \mathcal{R}_{in}$ model temporal relations between notes~\cite{karystinaios2023voice}.
$\mathcal{R}_{in}$ includes 4 types of relations: onset, during, follow, and silence, corresponding, respectively, to two notes starting at the same time, a note starting while the other is sounding, a note starting when the other ends, and a note starting after a time when no note is sounding. We also create the inverse edges for during, follows, and silence relations.
Each node corresponds to a vector of features: one of the 12 note pitch classes\footnote{We don't consider tonal pitch classes~\cite{pkspell} since they are not specified in MIDI files which we assume to be our input.} (C, C\#, D, etc.), the octave in $[1,\dots,7]$, the note duration, encoded as a float value $d \in [0,1]$ computed as the ratio of the note and bar durations, passed through a tanh function to limit its value and boost resolution for shorter notes, as proposed in~\cite{karystinaios2023voice}.
We don't consider grace notes in our system, and we remove them from the input notes.

\subsection{Output Graph}\label{subsec:output_graph}
The output graph $G_{out}=(V, E_{out}, \mathcal{R}_{out})$ has the same set $V$ of nodes as the input graph, but a staff number in $\{0,1\}$ is assigned to every node. There are two edge types in $E_{out}$: chord and voice, i.e. $\mathcal{R}_{out} = \{\text{"chord"}, \text{"voice"}\}$.

Voice edges~\cite{duane2009streaming,karystinaios2023voice} are an alternative in the literature 
to the more straightforward approach of predicting a voice number for every note; the usage of voice edges has the advantage of enabling a system to work with a non-specified number of voices, and avoiding the label imbalance problem for high voice numbers.
Voice edges are directed edges that connect consecutive notes (without considering rests) in the same voice. Formally, let $u_1,u_2 \in V$ be two notes in the same voice then $ (u_1, \text{"voice"}, u_2) \in E_{out}$ if and only if $\text{offset}(u_1) \leq \text{onset}(u_2)$ and $\nexists \ u_3 \in V$  within the same voice such that $\ \text{offset}(u_1) \leq \text{onset}(u_3) < \text{onset}(u_2)$. 

The previous definition also holds in our setting with homophonic voices. Let us extend the definition of \textit{chord} (a set of synchronous notes) to include the limit case of a single note. Two chords are consecutive if any two notes, respectively, from the first and second chords are consecutive. In the case of two consecutive chords with $m$ and $n$ notes in the same voice, there will be $m*n$ voice edges.

Chord edges are undirected and connect all notes that belong to the same chord without self-loops, so for a $n$-note chord, there are $n(n-1)$ edges.
They serve to unambiguously identify which notes together form a single chord. 

The same output graph can be created from an already properly engraved score. To obtain the graph we only need to draw the true voice edges between consecutive notes in the same voice within a bar and for chord edges we draw the chord ground truth between synchronous notes with the same voice number assignment. This graph can subsequently serve as the ground truth during training.

\subsection{Problem Simplification}\label{subsec:simplication}

In this section, we apply some obvious musical constraints to reduce computation and memory usage in our pipeline, without impacting the results. Let us first focus on \textit{chord edge} prediction. Given the simple constraint that all notes of a chord must start and end simultaneously, we can restrict the chord edge prediction process to only consider pairs of sychronous notes (same onset and offset values) as candidates. We do this by creating a set of \textit{chord edge candidates} $\Lambda_c$ which are calculated automatically and associated with our input graph. Only notes connected by such candidate edges will be considered in the chord prediction part of the model (see next section).


The same reasoning can be applied to the voice edges, by creating a set of \textit{voice edge candidates} $\Lambda_v$ such that $\forall u_1, u_2 \in V, \; (u_1, \text{"voice"}, u_2)\in \Lambda_v$ only when $\text{offset}(u_1) > \text{onset}(u_2)$. Another step can be taken towards reducing the number of candidates in the set $\Lambda_v$ by incorporating some musical engraving considerations.

The separation of notes in multiple voices does not have to be consistent in the whole score, but only within each bar, to produce the intended visual representation. There are no graphical elements that show if two notes in different bars are or are not in the same voice\footnote{This may change for cross-bar beamings, but they are very rarely used in standard music notation (there are no occurrences in our datasets) and therefore we do not consider them in this work.}.
Music engraving software does not force users to use consistent voices across bars. This can be often observed in digitized musical scores where music motives that belong to the same voice, are assigned different voices in different bars. Such observations have motivated projects such as the Symbolic Multitrack Contrapuntal Music Archive~\cite{aljanaki2021mcma} that explicitly encode a ``global'' voice number.

Since cross-bar consistency is not necessary for our goal of engraving (and is often wrongly annotated in our data) 
we limit the \textit{voice edge candidates} $\Lambda_v$ to contain only pairs of notes that occur within the same bar. This design choice is also reflected in our evaluation, i.e. we do not evaluate how the voices propagate across bars, but only within each bar. Note that this process is different from processing each bar independently since our network (detailed in the next section) considers music content across bars.

\subsection{Model}\label{subsec:model}

\begin{figure*}[tbp]
    \centering
    \includegraphics[width=0.9\textwidth]{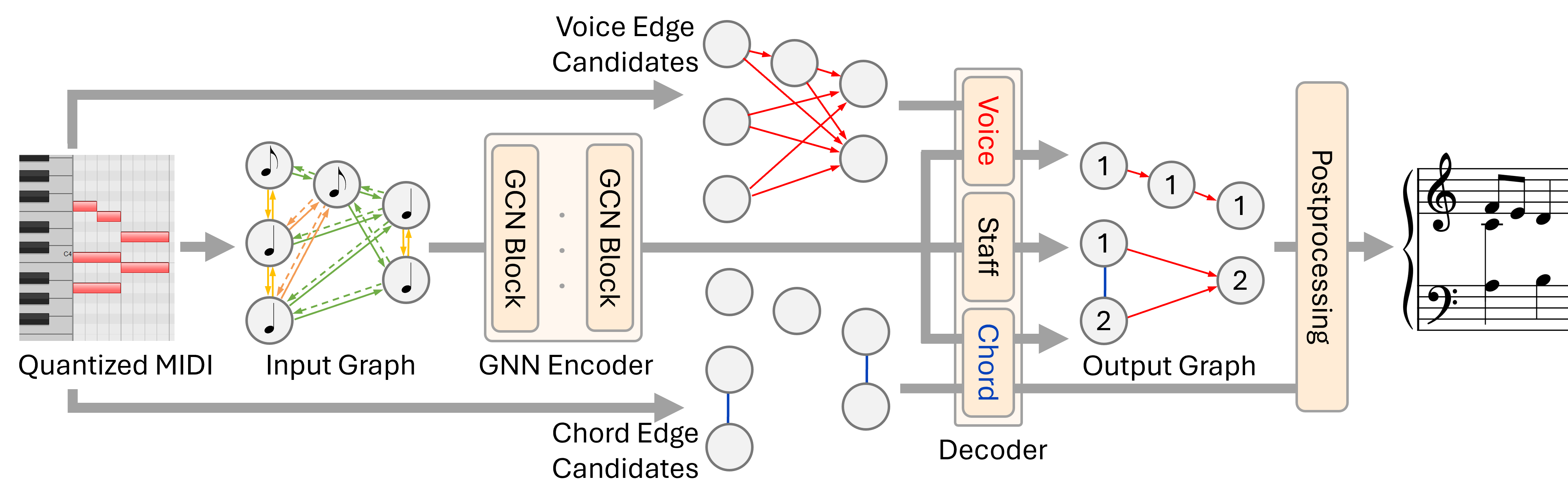}
    \caption{Our Architecture. For simplification, we display the output graph as having ``hard'' voice predictions, while these are probabilities over voice candidates.}
    \label{fig:model}
\end{figure*}
We design an end-to-end model (see Figure~\ref{fig:model}) that receives an input graph as described in Section~\ref{subsec:input_graph} and produces an output graph as in Section~\ref{subsec:output_graph}. The model is organized as an encoder--decoder architecture. 

The encoder receives an input graph created from a quantized MIDI score and passes it through three stacked Graph Convolutional Network (GCN) blocks to produce a node embedding for each note.
We use the heterogeneous version of the Sage convolutional block~\cite{hamilton2017representation} with a hidden size of 256; the update function for each node $u$ is described by:
\begin{align}
    \begin{aligned}
        \mathbf{h}_{\mathcal{N}(u)}^{(l+1)} &= \sum
        \left(\{\mathbf{h}_{v}^{l}, \forall v \in \mathcal{N}(u) \}\right)\\
        \mathbf{h}_{u}^{(l+1)} &= \sigma \left(\mathbf{W} \cdot \mathrm{concat}
        (\mathbf{h}_{u}^{l}, \mathbf{h}_{\mathcal{N}(u)}^{l+1}) \right)\\
    \end{aligned}
\end{align}
where $\mathcal{N}(u)$ are the neighbors of node $u$, $\sigma$ is a non-linear activation function, $\mathbf{W}$ is a learnable weight matrix.

The decoder consists of three parts that all use the same node embedding as input: i) a staff predictor; ii) a voice edge predictor; and iii) a chord clustering (i.e., a chord edge predictor).
The \textit{staff predictor} is a 2-layer Multi-Layer Perceptron (MLP) classifier that produces probabilities for each graph node (i.e., each note) to belong to the first or second staff.
The \textit{voice edge predictor} receives the embeddings of pairs of notes connected by edge candidates and produces a probability for each pair to be in the same voice. It works by concatenating the pairs of note embeddings and applying a 2-layer MLP. 
The final decoder part, \textit{chord clustering}, receives the embeddings of pairs of notes connected by chord edge candidates (i.e., pairs of synchronous notes) and produces the probability for a pair to be merged into a chord. This is achieved by computing the cosine similarity between the elements of the pair. This process forces the node embeddings created by the decoder to be similar to each other for notes of the same chord, which helps the voice predictor produce consistent voice edge probabilities for notes of the same chord. We apply a threshold to pass from probabilities to decisions on which notes to cluster.


The complete model contains $\sim 3$M parameters and we train it end-to-end with the (unweighted) sum of three \textit{Binary Cross Entropy} loss functions, one for each task.



\subsection{Postprocessing}


A straightforward approach to deciding whether to connect two notes with a voice edge would be to threshold the predicted voice edge probabilities. However, even when using edge and chord candidates, we could still produce three kinds of invalid output: (1) multiple voices merging into one voice, (2) one voice splitting into multiple voices, and (3) notes in the same chord that are not in the same voice. To eliminate these issues, we add a postprocessing phase that accompanies our model and guarantees a valid output according to music engraving rules. 

The first step, which we call \textit{chord pooling}, merges all nodes that belong to the same chord to a single new "virtual node". This is done by looking for the connected components considering only chord edges in the output graph, then \textit{pooling} in a single node all original nodes in each connected component, creating a new node which has as incoming and outgoing voice edges all edges entering and exiting the original nodes, respectively. If multiple edges collapse in one edge (e.g. in the case of two consecutive chords in the same voice), the new edge has a probability that is the average of the corresponding edge probabilities. 
    

After the first step, we are left with monophonic streams, which could still exhibit problems (1) and (2). We can solve this with the technique proposed in~\cite{karystinaios2023voice} for monophonic voices, i.e. by framing the voice assignment problem as a linear assignment problem~\cite{burkard1999linear} over the adjacency matrix obtained by the updated edge candidates $\Lambda'_v$. We follow the linear assignment step by unpooling or unmerging the nodes that were previously pooled, in this way, obtaining the original nodes again. During unpooling, the incoming edges and outgoing edges of the "virtual nodes" are reassigned to each original node, thus resolving problem (3).

\begin{figure}[t]
    \centering
    \includegraphics[width=\columnwidth]{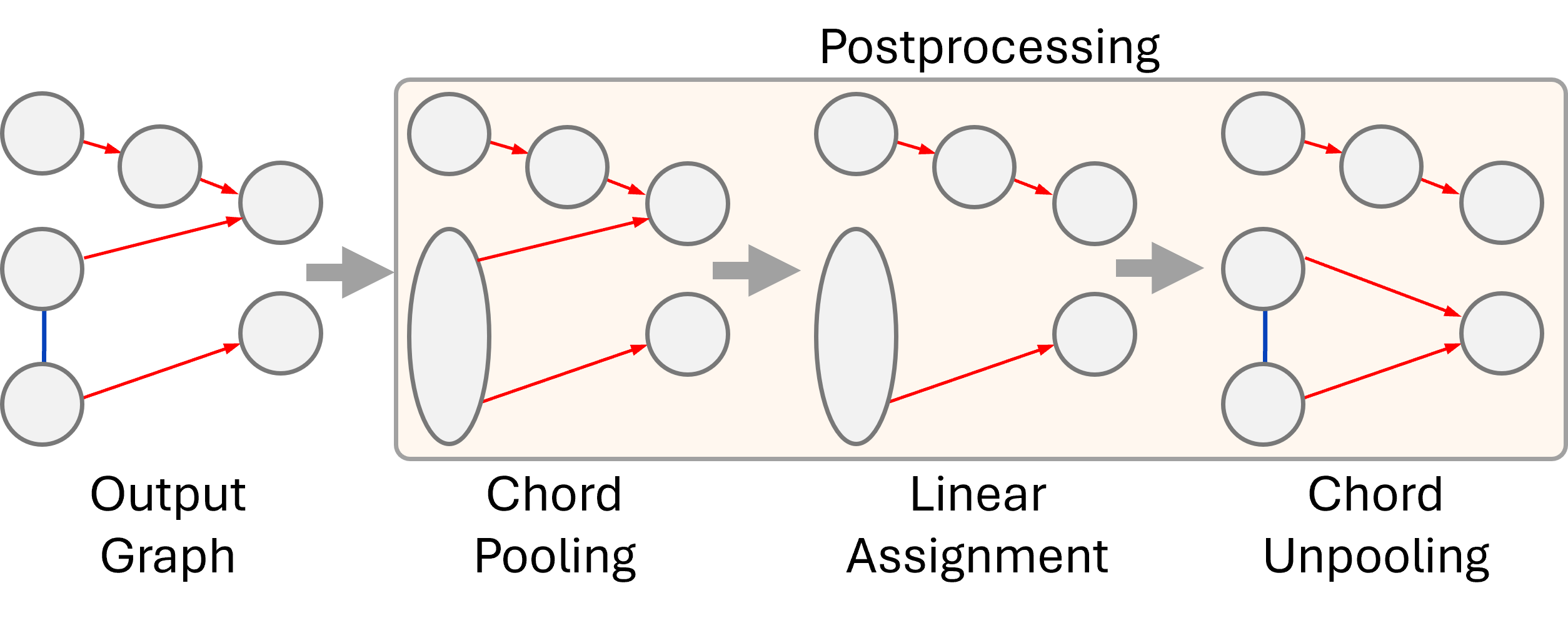}
    \caption{Output graph postprocessing. We do not display the predicted staff labels.} 
    \label{fig:postp}
\end{figure}

The complete postprocessing method is depicted in Figure~\ref{fig:postp}. It is worth noting that the staff labels are not considered during the postprocessing phase, and we copy them unchanged to the postprocessed output graph.


\subsection{Evaluation}\label{subsec:eval}
We evaluate the predicted voice assignments with the metric proposed by Hiramatsu et al.~\cite{hiramatsu2021joint}, which formalizes the metric of McLeod and Steedman~\cite{McLeod:metric}. 
This is a version of the F1-score for voice separation~\cite{duane2009streaming} which is adapted to work on homophonic voices, by reducing the importance of notes if they are part of a chord. This is important since chords create many voice edges (e.g., two 4-note chords in the same voice are connected by 16 edges), which could potentially overshadow the importance of edges in monophonic voices (or voices with fewer/smaller chords).

Formally, the homophonic voice F1-score $F1$ is calculated as:
\begin{equation}
    \begin{gathered}       
    P = \frac{\sum_{i < j} a_{ij} \hat{a}_{ij} / \hat{w}_i}{\sum_{i < j} \hat{a}_{ij} / \hat{w}_i} , \;\;\;
    R = \frac{\sum_{i < j} a_{ij} \hat{a}_{ij} / w_i}{\sum_{i < j} a_{ij} / w_i} \\
    F1 = \frac{2 P R}{P + R}    
    \end{gathered}
\end{equation}

where $i<j$, in the sum, considers all pair of notes $i,j$ such that $\text{offset}(i)<\text{onset}(j)$; $a_{ij}$, $\hat{a}_{ij}$ are equal to 1 or 0 if a voice edge exists or not in the ground truth and predictions, respectively; and $w_i$ and $\hat{w}_i$ are the number of notes that are chorded together with the note $i$ in the ground truth and predictions, respectively. Unlike~\cite{hiramatsu2021joint}, we consider only notes $j$ in the same bar of $i$, for the reasons presented in Section~\ref{subsec:simplication}.
We evaluate the staff prediction part of our model with binary accuracy, and we assess chord prediction with the F1 score computed on the chord edges.


\begin{table*}[!htbp]
  \centering
    \begin{tabular}{lllllll} 
    \toprule
     \multirow{2}{*}{Dataset} &
      \multicolumn{3}{c}{\textbf{J-pop Dataset}}  &      
      \multicolumn{3}{c}{\textbf{DCML Romantic Corpus}} \\
      \cmidrule(lr){2-4} 
      \cmidrule(lr){5-7}
         
                   & Staff Acc       & Chord F1                   & Voice F1                & Staff Acc        & Chord F1        & Voice F1    \\
    \midrule
    Baseline       & $89.9$                & $86.9 $                   & $85.4$                 & $80.7$           & $65.2$           & $78.2$         \\  
    Shibata et al.~\cite{shibata2021non} & $92.8$                & -                         & $92.2$                 & $88.5$           & -                & $84.9$         \\ 
    \midrule
    GNN wo Chord wo Post           & $\textbf{96.5}\pm 0.1$ & -                        & $95.2\pm 1.9$        & $\textbf{91.5}\pm 0.1$  & -                       & $87.2\pm 3.3$         \\
    GNN wo Post       & $96.3\pm 0.1$          & $\mathbf{94.9}\pm 0.1$   &$95.7\pm 0.4$& $91.0\pm 0.1$           & $\textbf{79.5}\pm 0.4$  & $88.9\pm 0.4$\\
    GNN & $96.3\pm 0.1$          & $\mathbf{94.9}\pm 0.1$   &$\mathbf{96.6}\pm 0.1$& $91.0\pm 0.1$           & $\textbf{79.5}\pm 0.4$  & $\mathbf{89.9}\pm 0.2$\\
    \bottomrule
    
  \end{tabular}
  \caption{
  Metrics for our the J-Pop and DCML test sets.
  ``GNN'' denotes our method, without postprocessing (``GNN wo Post''), and without both postprocessing and chord prediction parts (``GNN wo Chord wo Post'').
  All GNN model runs are repeated 5 times: $\pm$ refers to the standard deviation of results across runs.}
  \label{tab:main_results}
\end{table*}


\subsection{From Network Prediction to Readable Output}
The computation of voice and staff numbers is sufficient for the system evaluation, but not for producing a usable tool, which we are interested in in this paper.
The missing step, to be described in this section, is the integration of the network predictions into a readable musical score. To achieve this integration we need to undertake two essential steps: beam together notes within the same voice, and infill rests to "fill holes" within each voice.

For the first step, we proceed according to the rules of engraving~\cite{gould2016behind}. We beam two consecutive notes (or chords) in the same voices if their duration is less than a quarter note (excluding ties) unless they belong to different beats. Following the music notation convention we consider the compound time signatures, i.e., $\frac{6}{x}$,$\frac{9}{x}$,$\frac{12}{x}$ to have, respectively, 2,3, and 4 beats. When confronted with tied notes, the algorithm prioritizes producing notations with the fewest number of notes, an heuristic with promotes easier-to-read notation~\cite{foscarin2019modeling}.

The second step consists of introducing rests so that each voice fills the entire bar and can be correctly displayed. 
Some rests could be set as invisible to improve the graphical output when their presence and duration are easy to assume from other score elements, but we display all of them for simplicity. 
As for the notes, we choose the rest types (with eventual dots) to minimize the number of rests in the score.

The two steps described above cover common cases and produce a complete score in MEI format~\cite{roland2002music}. However, the score export is still a prototype, since developing one that is robust against all corner cases is an extremely complex task, and is outside the scope of this paper. Since score output problems may obscure the output of our system, we also develop a graph visualization tool. Both the input and output graphs (including the candidate edges) can be displayed on top of the musical score in an interactive web-based interface based on Verovio~\cite{Pugin2014VerovioAL}. Some examples of the output graph visualization are in Figure~\ref{fig:comparison}.

\section{Experiments}\label{sec:experiments}

We train our model with the ADAM optimizer with a learning rate of $0.001$ and a weight decay of $5*10^{-4}$ for 100 epochs.
For a quantitative evaluation, we compare our results with those of a baseline algorithm and the method proposed by Shibata et al.~\cite{shibata2021non}, on two rather diverse datasets. 

Our baseline algorithm assigns all notes under C4 (middle C) to the second staff and the rest to the first. Then it groups all synchronous notes (per staff) as chords. 
Finally, it uses the time and pitch distances between the candidate pairs of notes as weights to be minimized during the linear assignment process (the same as we use in our postprocessing) which creates the voice edges.

\subsection{Datasets}\label{subsec:datasets}
We use two piano datasets of different styles and difficulties to evaluate our system under diverse conditions. The ability to handle complex corner cases should not reduce the performance on easier (and more common) pieces.

    The \textit{J-Pop} dataset contains pop piano scores introduced by~\cite{shibata2021non}. Most of the scores consist of accompaniment chords on the lower staff and some simple melodic lines on the upper staff. The dataset contains 811 scores; we randomly sampled 159 (roughly 20\%) of these for testing and used the rest for training and validation.

The \textit{DCML Romantic Corpus} is more challenging. It was created by~\cite{dcml_piano_corpus} and contains piano pieces from the 17th to 20th centuries with some virtuosic quality. It includes characteristics such as cross-staff beaming, a high number of voices, challenging voicing, etc. Similarly to the pop dataset we randomly sample 77 out of the 393 scores (approx.~20\%) and use the rest for training and validation.

The \textit{J-Pop} dataset is available in MusicXML format, while the \textit{DCML Romantic Corpus} is in Musescore file format. We use Musescore to convert DCML files to MusicXML and load them with the Python library Partitura~\cite{partitura_mec} to extract the note list.

\subsection{Results}\label{subsec:results}
Our model aims to be generic across a variety of music, therefore we train a single model on the joined training set of pop and classical scores, not two individual ones. The rules that govern the handling of voices may be fundamentally different in the two datasets, but we assign to the model the task of handling these differences. This approach ensures better future scalability on bigger and more diverse datasets. We compute the metrics separately on the test part of our two datasets.

Table~\ref{tab:main_results} reports results for three versions of our graph-based model: the complete model from Section~\ref{sec:Methodology}, a variant without postprocessing, and a variant without chord prediction and postprocessing (our postprocessing technique cannot be run without the chord prediction task, since it pools nodes that belong to the same predicted chord).
The method of Shibata requires the specification of the number of voices per staff. For compactness, we report only the results with one voice per staff (2 voices total); the results degrade by increasing the number of voices.

Our results show that even our system without pooling and without postprocessing obtains consistently better results than both Shibata et al.~\cite{shibata2021non} and our baseline.
Interestingly, the chord prediction task improves the Voice F1 results even when the post-processing is not used; this confirms the benefits of multi-task training, and of enforcing notes of the same chord to have similar representations in the hidden space, with cosine similarity, to predict coherent voice edges. However, we observe a reduction in staff accuracy, probably for the same reason, since the same hidden representation is also used to predict chords, making it harder (though not impossible) to split notes of the same chord in different staves.
When the full system is used, there are further improvements in Voice F1.


We are also evaluate our system on the bar-level and study performances for music excerpts of varying difficulties. We compute the voice F1 score for each bar and average them based on the number of voices in the ground truth. We compare with Shibata et al.~\cite{shibata2021non} with 1 \& 2 voices per staff (vps). Table~\ref{tab:measurewise} shows the results for the DCML Romantic Corpus. Both our model and \cite{shibata2021non} perform best with 2 voices, the most common number in our dataset. Interestingly, Shibata et al. approach with 2 vps never outperforms vps 1, not even when the target number of voices is 3 or 4, a situation that vps 1 cannot handle. This can be explained by the fact that Shibata et al. parameters were tuned on a simpler dataset, and accepting more voices creates more errors than benefits. Setting vps $>2$ consistently degraded the performances, probably also for similar reasons.

\begin{table}[htbp]
\centering
\begin{tabular}{ccccc}
\toprule
\#Voices & \#Bars & \textbf{GNN} & \textbf{\cite{shibata2021non} 1vps} & \textbf{\cite{shibata2021non} 2vps}  \\ 
\cmidrule(lr){1-2}
\cmidrule(lr){3-5}
1 &  322 & 96.6 & 88.3 & 87.9  \\
2 & 4576 &  94.1 & 89.3 & 88.1  \\
3 & 2464 &  89.0 & 84.2 & 81.5  \\
4 & 719 & 81.6 & 80.5 & 75.1  \\
5 & 99 & 81.6 & 76.7 & 73.7  \\
6  & 17 & 78.4 & 68.9 & 61.6 \\
\bottomrule
\end{tabular}
\caption{Voice F1 score aggregated by bars with the same number of voices in the ground truth, on the DCML Corpus. Shibata et al.~\cite{shibata2021non} is used with 1 and 2 voices per staff (vps).}
\label{tab:measurewise}
\end{table}

\begin{figure}[htp]
    \centering
    \includegraphics[width=\columnwidth]{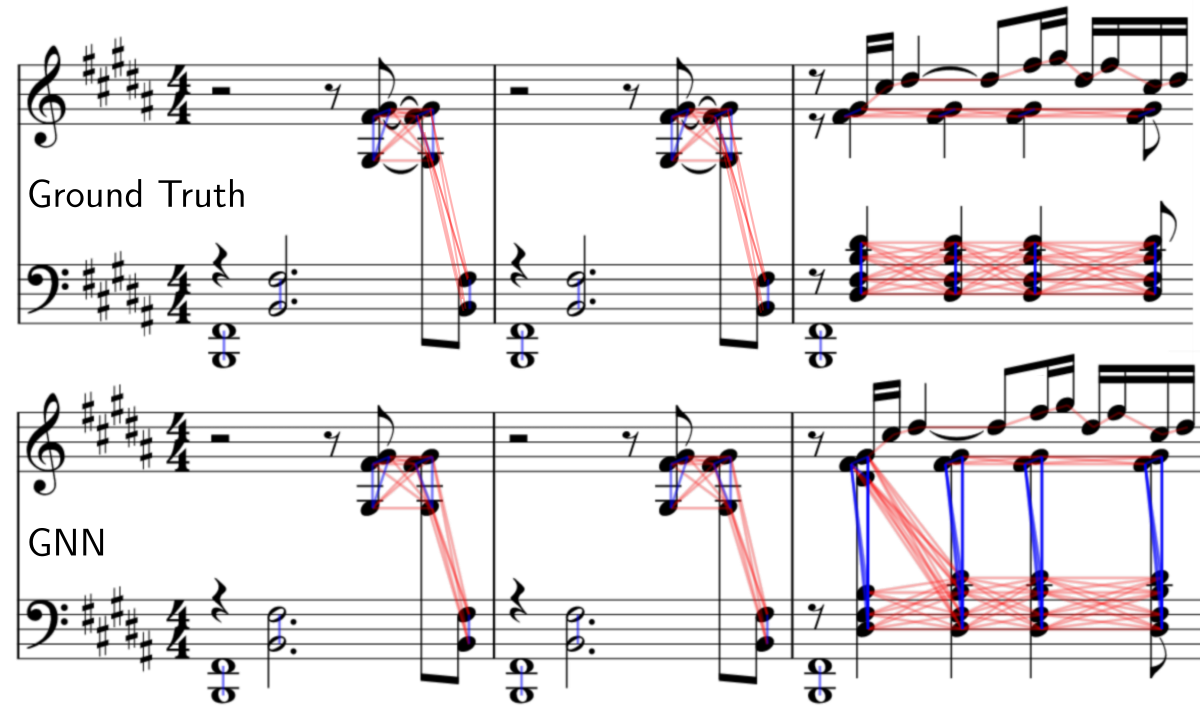}
    \caption{Comparison of voice and staff assignment between the original score (Ground Truth) and our method (GNN) on the first bars of C. Debussy's Estampes-Pagodes. Voice edges are drawn in red and chord edges in blue.
    }
    \label{fig:comparison}
\end{figure}

\subsection{Qualitative Analysis}\label{subsec:q_analysis}




Let us take a closer look into the predictions of our deep-learning approach (GNN) on the excerpt of Figure~\ref{fig:comparison} produced by our visualization tool. Our approach captures correctly the cross-staff voice in the first two bars, while such a situation causes performance degradation for all other voice separation approaches that don't support it. We observe some disagreements with the original score in Measure 3: our model predicts a single chord (instead of splitting across the staff) containing all the synchronous syncopated quarter notes, and also mispredicts the staff for the first D\#4 note. A more in-depth study of why this happens is not trivial, as neural networks are not interpretable. This is a drawback compared to heuristic-based separation techniques.

Synchronous notes with the same pitch 
are problematic. Our system can predict different voices for these notes, while Shibata et al. always predict them as a chord in the same voice, and this reduces the performances for pieces that contain a lot of them, like Schumann Kinderszenen Op.15. For fairness, we should note that we should expect the output of a music transcription system to only contain one of these notes, instead of multiple like in our current input. An enhancement of our system would then be able to receive a single note as input, assign multiple voices to it (with multiple incoming and outgoing edges) and then split it into multiple notes.
Another current limitation of our system is the missing support for grace notes, which in the actual version are ignored and removed from the input.

\section{Conclusion and Future Work}\label{sec:conclusion}

This paper presented a novel graph-based method for homophonic voice separation and staff prediction in symbolic piano music. 
Our experiments highlight our system's effectiveness compared to previous approaches. Notably, we obtained consistent improvements over two datasets of different styles with a single model.

Future work will focus on integrating grace notes and the possibility of multiple voices converging on a single note. We aim to create a framework that produces complete engravings from quantized MIDI, including the prediction of clef changes, beams, pitch spelling, and key signatures.

\section{Acknowledgements}
This work is supported by the European Research Council (ERC) under the EU's Horizon 2020 research \& innovation programme, grant agreement No.\ 101019375 (\textit{Whither Music?}), and the Federal State of Upper Austria (LIT AI Lab).

\bibliography{biblio}
\end{document}